%% file: Sequential-attractors-SIAM-DS-Web-oct20.tex
\documentclass[11pt]{article}

\input{packages-v2}

\usepackage{helvet}
\usepackage{graphicx,amssymb,amsmath,amsfonts,amsthm,color,url, placeins, enumitem}
\usepackage{wrapfig}
\usepackage{tabstackengine}
\usepackage{multirow,ctable} 
\usepackage{makecell}

\renewcommand\section{\@startsection{section}{1}{\z@}%
                                   {-3.0ex \@plus -1ex \@minus -.2ex}%
                                   {1.5ex \@plus.2ex}%
                                   {\normalfont\sffamily\large\bfseries}}
\renewcommand\subsection{\@startsection{subsection}{2}{\z@}%
                                     {-2.75ex\@plus -1ex \@minus -.2ex}%
                                     {1.5ex \@plus .2ex}%
                                   {\normalfont\sffamily\large}}
\renewcommand\subsubsection{\@startsection{subsubsection}{3}{\z@}%
                                     {-2.75ex\@plus -1ex \@minus -.2ex}%
                                     {1.5ex \@plus .2ex}%
                                   {\normalfont\sffamily\large}}

\newcommand{\od}{\stackrel{\mbox {\tiny {def}}}{=}}

\def\d{\mathrm{d}}

\def\max{\mathrm{max}}

\def\supp{\operatorname{supp}}

\def\od{\stackrel{\mathrm{def}}{=}}

\def\supp{\operatorname{supp}}

\def\FP{\operatorname{FP}}

\usepackage{color}
\definecolor{cherry}{rgb}{0.9,.1,.2}

\begin{document}

\noindent {\large \bf Sequence generation in inhibition-dominated neural networks}\\
\noindent Caitlyn Parmelee, Juliana Londono Alvarez, Carina Curto*, Katherine Morrison* \\
{\footnotesize * equal contribution}\\

Sequences of neural activity arise in many brain areas, including cortex \cite{Luczak-PNAS, Yuste-cortex-sequences, Yuste-CPG}, hippocampus \cite{Stark-PNAS, Eva-science, JNeuro}, and central pattern generator circuits that underlie rhythmic behaviors like locomotion \cite{Marder-CPG, CPG-review}.  Such sequences are examples of emergent or \emph{internally-generated activity}: neural activity that is shaped primarily by the structure of a recurrent network rather than inherited from a changing external input.  A fundamental question is to understand how a network's connectivity shapes neural activity.  In particular, what types of network architectures give rise to sequential activity?  

While the architectures underlying sequence generation vary considerably, a common feature is an abundance of inhibition. Inhibition-dominated networks exhibit emergent sequences even in the absence of an obvious chain-like architecture, such as a synfire chain \cite{Kopell2, Bazhenov-hipp-ripples}. Here, we provide an overview of results from \cite{Parmelee2022} on network architectures that produce sequential dynamics in a special family of inhibition-dominated networks known as combinatorial threshold-linear networks (CTLNs).  

\vspace{-10pt}
\paragraph{Emergent sequences from CTLNs\\}

Combinatorial threshold-linear networks (CTLNs) are a special family of threshold-linear networks (TLNs), whose dynamics are prescribed by the equations:
\begin{equation}\label{eq:dynamics}
\dfrac{dx_i}{dt} = -x_i + \left[\sum_{j=1}^n W_{ij}x_j+\theta_i \right]_+, \quad i = 1,\ldots,n.
\vspace{-.04in}
\end{equation}
Here, $x_1(t),\ldots,x_n(t)$ represent the firing rate of $n$ recurrently-connected neurons, $[\cdot]_+ = \max\{0,\cdot\}$ is the threshold nonlinearity, and the connection strength matrix $W = W(G,\varepsilon,\delta)$ is determined by a simple\footnote{A graph is \emph{simple} if it does not have self-loops or multiple edges (in the same direction) between a pair of nodes.}
 directed graph $G$ as follows:
\begin{equation} \label{eq:binary-synapse}
W_{ij} = \left\{\begin{array}{ll} \phantom{-}0 & \text{ if } i = j, \\ -1 + \varepsilon & \text{ if } j \rightarrow i \text{ in } G,\\ -1 -\delta & \text{ if } j \not\rightarrow i \text{ in } G. \end{array}\right. \quad \quad \quad \quad
\end{equation} 
We require the parameters to satisfy $\delta >0$, and $0 < \varepsilon < \frac{\delta}{\delta+1}$, and typically we take the external input to be constant, $\theta_i=\theta>0$, to ensure the dynamics are internally generated.  Note that the upper bound on $\varepsilon$ implies $\varepsilon < 1$, and so the $W$ matrix is always effectively inhibitory.  We think of this as modeling the activity of excitatory neurons in a sea of inhibition: an edge indicates an excitatory connection in the presence of background inhibition, while the absence of an edge indicates no excitatory connection. The graph $G$ thus captures the effective pattern of weak and strong inhibition in the network.  

\begin{figure}[!ht]
\begin{center}
\includegraphics[width=.8\textwidth]{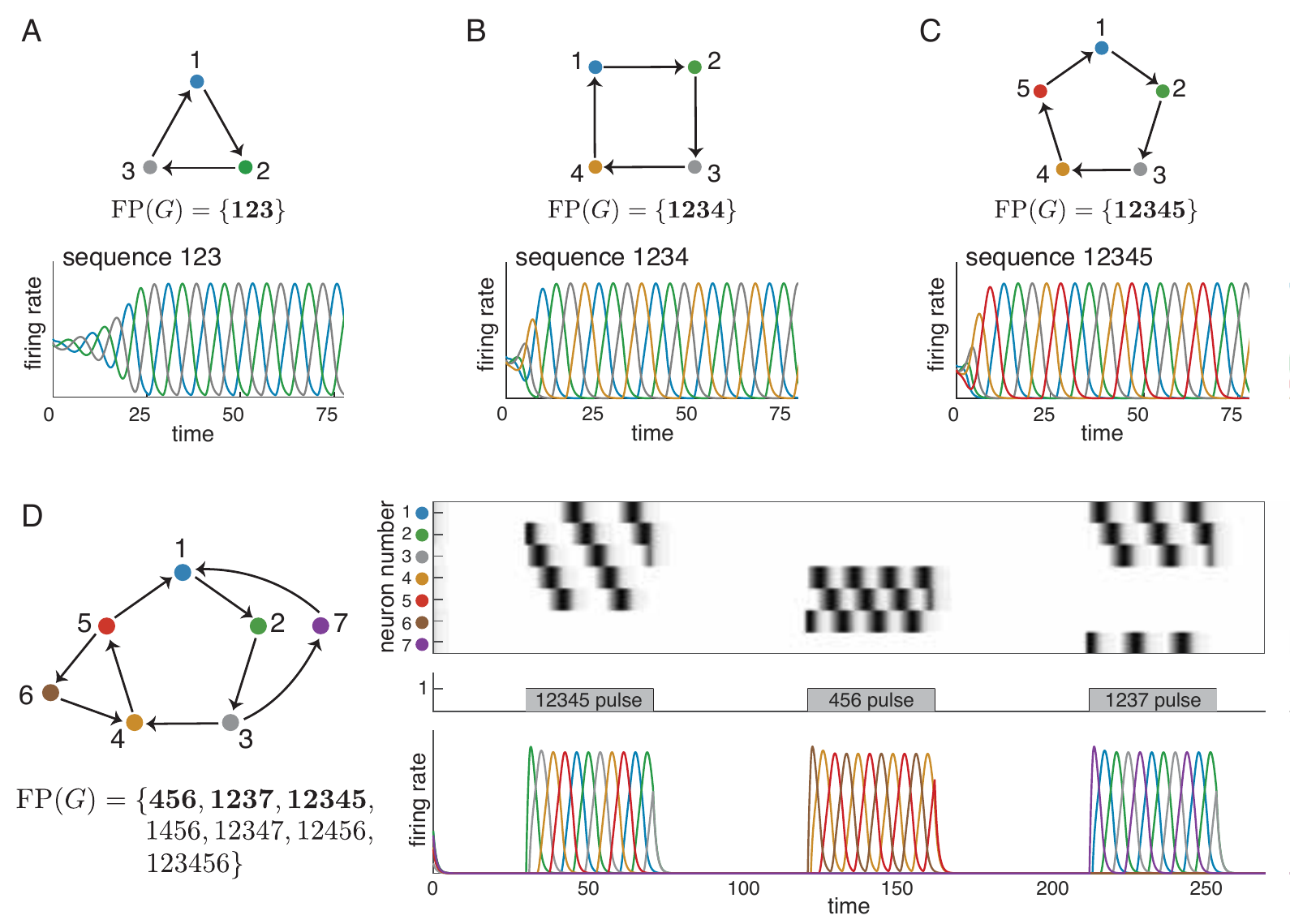}
\end{center}
\vspace{-10pt}
\caption{{\bf Sequential attractors from simple cycles.} (A-C) CTLNs corresponding to a $3$-cycle, a $4$-cycle, and a $5$-cycle each produce a limit cycle where the neurons reach their peak activations in the expected sequence. Colored curves correspond to solutions $x_i(t)$ for matching node $i$ in the graph. 
(D) Attractors corresponding to the embedded $3$-cycle, $4$-cycle, and $5$-cycle of the network are transiently activated to produce sequences matching those of the isolated cycle networks in A-C.  For each network in A-D, $\FP(G)$ is shown, with the minimal fixed points in bold. Note that we say a fixed point is \emph{minimal} if its support is minimal under inclusion within $\FP(G)$.  To simplify notation for $\FP(G)$, we denote a subset $\{i_1, \ldots, i_k\}$ by $i_1\cdots i_k$.  For example, $12345$ denotes the set $\{1,2,3,4,5\}$. Unless otherwise noted, all simulations have CTLN parameters $\varepsilon=0.25, \delta = 0.5,$ and $\theta =1$.}
\label{fig:sequences-setup}
\vspace{-8pt}
\end{figure}

The simplest architectures that give rise to sequences in CTLNs are graphs that are cycles.  These networks produce limit cycles in which the neurons reach their peak activation in cyclic order.  
For example, in Figure~\ref{fig:sequences-setup}A-C, we see such sequential limit cycles when the graph $G$ is a $3$-cycle, $4$-cycle, or a $5$-cycle.
Interestingly, these simple cycles can also be embedded in a larger network, as in Figure~\ref{fig:sequences-setup}D, and still produce sequences that match those of the subnetwork in isolation. Each of the sequences from panels A-C may be transiently activated in the network in panel D by a time-dependent external drive $\theta_i(t)$ consisting of pulses that are concentrated on different subnetworks. A single simulation is shown, with localized pulses activating the $5$-cycle, the $3$-cycle, and finally the $4$-cycle. Although these cycles overlap, each pulse activates a sequence involving only the neurons in the stimulated subnetwork. Depending on the duration of the pulse, the sequence may play only once or repeat two or more times.

What other architectures, beyond simple cycles, produce sequential dynamics? Is there a key feature that we can generalize from cycles that will allow us to build more interesting architectures supporting sequential activity? To better understand how graph connectivity shapes dynamics, we focus our attention on the relationship between connectivity and fixed points of the dynamics.

\paragraph{Fixed points of CTLNs\\}

One of the most striking features of CTLNs is the strong connection between dynamic attractors and \emph{unstable} fixed points \cite{book-chapter, rule-of-thumb}.  (Of course, stable fixed points are also attractors of the network, but these are static.)  Recall that a fixed point $x^*$ of a CTLN is a solution that satisfies $dx_i/dt|_{x=x^*} = 0$ for each $i \in [n]$. The  {\it support} of a fixed point is the subset of active neurons, $\supp{x} \od \{i \mid x^*_i>0\}$. For a given nondegenerate TLN, there can be at most one fixed point per support, and it is straightforward to recover the actual fixed point values from knowledge of the support.  Thus, it suffices to label all the fixed points of a network by their support, $\sigma = \supp{x^*} \subseteq [n],$ where $[n] \od \{1, \ldots, n\}$.  We denote the collection of fixed point supports by 
\vspace{-.05in}
\[\FP(G) = \FP(G, \varepsilon, \delta)\od \{\sigma \subseteq [n] ~|~  \sigma \text{ is a fixed point support of } W(G,\varepsilon,\delta) \}.\]
For example, in Figure~\ref{fig:sequences-setup}A, $123$ is the only fixed point support of the $3$-cycle. In contrast, the network in Figure~\ref{fig:sequences-setup}D supports several fixed points. Can these fixed point supports give insight into the dynamics? 

It turns out that the \emph{minimal} fixed points of the network in Figure~\ref{fig:sequences-setup}D reflect the subsets of neurons that are active in each of the attractors.  This phenomenon has been consistently observed in simulations; in \cite{rule-of-thumb} it was shown that there is a close correspondence between certain minimal fixed points in $\FP(G)$, known as \emph{core motifs}, and the attractors of a network. Additionally, prior work proved a series of {\it graph rules} that can be used to determine fixed points of a CTLN by analyzing the structure of the graph $G$ \cite{fp-paper, stable-fp-paper}. These rules directly connect the network connectivity to $\FP(G)$, and are all independent of the choice of parameters $\varepsilon, \delta,$ and $\theta$.  Here, we investigate families of network architectures that give strong constraints on $\FP(G)$, and for which this fixed point structure is predictive of sequential attractors of the networks.


\vspace{-10pt}
\paragraph{What other architectures can support sequential attractors?\\} 
Simple cycles were the most obvious candidate to produce sequential attractors. But what other architectures can support sequential attractors? Here, we investigate four architectures that generalize different features of simple cycles: cyclic unions, directional cycles, simply-embedded partitions, and simply-embedded directional cycles. A common feature of all these architectures is that the neurons of the network are partitioned into components $\tau_1,\ldots,\tau_N$, organized in a cyclic manner, whose disjoint union equals the full set of neurons $[n] \od \{1,\ldots,n\}$. The induced subgraphs $G|_{\tau_i}$ are called component subgraphs.  In \cite{Parmelee2022} we prove a series of theorems about these architectures, connecting the fixed points of a graph $G$ to the fixed points of the component subgraphs $G|_{\tau_i}$.  In the following, we give an overview of these architectures and summarize the key results proven about them in \cite{Parmelee2022}.

\begin{figure}[!h]
\begin{center}
\includegraphics[width=.77\textwidth]{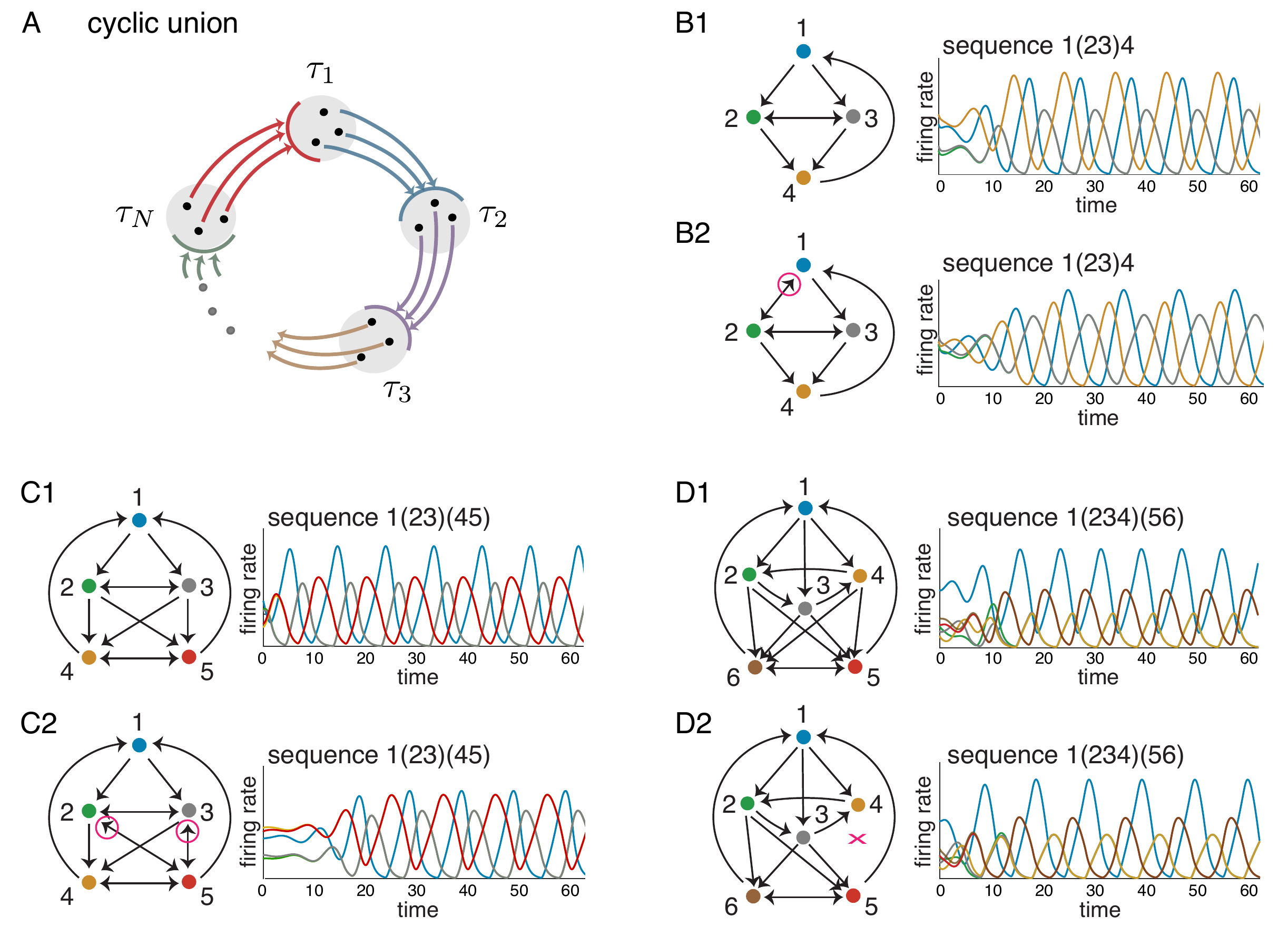}
\caption{{\bf Cyclic unions and related variations.}  (A) A cyclic union has component subgraphs with subsets of nodes $\tau_1, \ldots, \tau_N$, organized in a cyclic manner. (B1,C1,D1) Three cyclic unions with firing rate plots showing solutions to a corresponding CTLN.  Above each solution the associated sequence of firing rate peaks is given, with synchronously firing neurons denoted by parentheses.  (B2,C2,D2) These graphs are all variations on the cyclic unions above them, with some edges added or dropped (highlighted in magenta).  Solutions of the corresponding CLTNs qualitatively match the solutions of the corresponding cyclic unions. In each case, the sequence is identical.}
\label{fig:cycu-generalizations}
\end{center}
\vspace{-.25in}
\end{figure}

\vspace{-12pt}
\paragraph{\it Cyclic unions.}
The most straightforward generalization of a cycle is the {\it cyclic union}, an architecture first introduced in \cite{fp-paper}. Given a set of component subgraphs $G|_{\tau_1}, \ldots, G|_{\tau_N},$ on subsets of nodes $\tau_1, \ldots, \tau_N$, the \emph{cyclic union} is constructed by connecting these subgraphs in a cyclic fashion so that there are edges forward from every node in $\tau_i$ to every node in $\tau_{i+1}$ (cyclically identifying $\tau_N$ with $\tau_0$), and there are no other edges between components (see Figure~\ref{fig:cycu-generalizations}A).

The top graphs in Figure~\ref{fig:cycu-generalizations}B-D are examples of cyclic unions with three components.  All the nodes at a given height comprise a $\tau_i$ component.  Next to each graph is a solution to a corresponding CTLN, which is a global attractor of the network. Notice that the activity traverses the components in cyclic order.   Cyclic unions are particularly well-behaved architectures where the fixed point supports can be fully characterized in terms of those of the components.  Specifically, the fixed points of a cyclic union $G$ are precisely the unions of supports of the component subgraphs, exactly one per component \cite{fp-paper}, that is, 
\vspace{-.1in}
$$\sigma \in \FP(G) \quad \Leftrightarrow \quad \sigma \cap \tau_i \in \FP(G|_{\tau_i})~~\text{ for all } i \in [N].$$
\vspace{-.15in}

The bottom graphs in Figure~\ref{fig:cycu-generalizations}B-D have very similar dynamics to the ones above them, but do not have the perfect cyclic union structure.  Despite deviations from the cyclic union architecture, these graphs produce sequential dynamics that similarly traverse the components in cyclic order. In fact, they are examples of a more general class of architecture: directional cycles.

\vspace{-12pt}
\paragraph{\it Directional cycles.}
In a cyclic union, if we restrict to the subnetwork consisting of a pair of consecutive components, $G|_{\tau_i \cup \tau_{i+1}}$, we find that activity initialized on $\tau_i$ flows forward and ends up concentrated on $\tau_{i+1}$.  Moreover, the fixed points of $G|_{\tau_i \cup \tau_{i+1}}$ are all confined to live in $\tau_{i+1}$, and so the concentration of neural activity coincides with the subnetwork supporting the fixed points. We say that a graph is \emph{directional} whenever its fixed points are confined to a proper subnetwork. In simulations, we have seen that directional graphs have the desired directionality of neural activity, so that activity flows to the subnetwork supporting the fixed points.  

We can chain together directional graphs in a cyclic manner to produce \emph{directional cycles}. We predict that these graphs will have a cyclic flow to their dynamics, hitting each $\tau_i$ component in cyclic order. While we have not been able to explicitly prove this property of the dynamics, we can prove that all the fixed point supports have cyclic structure (Theorem 1.2 in \cite{Parmelee2022}). However, unlike with cyclic unions, for directional cycles we do not necessarily have the property that fixed points $\sigma$ of the full network restrict to fixed points $\sigma_i\od \sigma \cap \tau_i$ of the component subnetworks $G|_{\tau_i}$. The key structural feature of cyclic unions that guarantees this property of the fixed points is what we call we call a \emph{simply-embedded partition}. 

\vspace{-12pt}
\paragraph{\it Simply-embedded partitions.}

Given a graph $G$, a partition of its nodes $\{\tau_1|\cdots|\tau_N\}$ is called \emph{simply-embedded} if all nodes in a component $\tau_i$ are treated identically by each node outside that component. For example, the pair of graphs in Figure~\ref{fig:cycu-generalizations}C have $\{1\, |\, 2,3\, |\, 4,5\}$ as a simply-embedded partition: in each graph, nodes 2 and 3 receive identical inputs from 1, 4 and 5 and nodes 4 and 5 receive identical inputs from 1, 2 and 3. It turns out that this simply-embedded partition structure is sufficient to guarantee that all the fixed points of $G$ restrict to fixed points of the component subgraphs. This means that the fixed points of the components provide a kind of ``menu'' from which the fixed points of $G$ are made: each fixed point of the full network has support that is the union of component fixed point supports, at most one per component (Theorem 1.4 in \cite{Parmelee2022}). 

\vspace{-12pt}
\paragraph{\it Simply-embedded directional cycles.}
While the simply-embedded partition generalizes one key property of $\FP(G)$ from cyclic unions, it does not guarantee that every fixed point intersects every component, nor does it guarantee the cyclic flow of the dynamics through the components.  But combining the previous two constructions, we immediately see that a \emph{simply-embedded directional cycle} will have the desired fixed point properties while maintaining cyclic dynamics. In particular, every fixed point support of $G$ is a union of (nonempty) component fixed point supports, exactly one per component (Theorem 1.5 in \cite{Parmelee2022}).

\begin{figure}[!ht]
\begin{center}
\includegraphics[width=\textwidth]{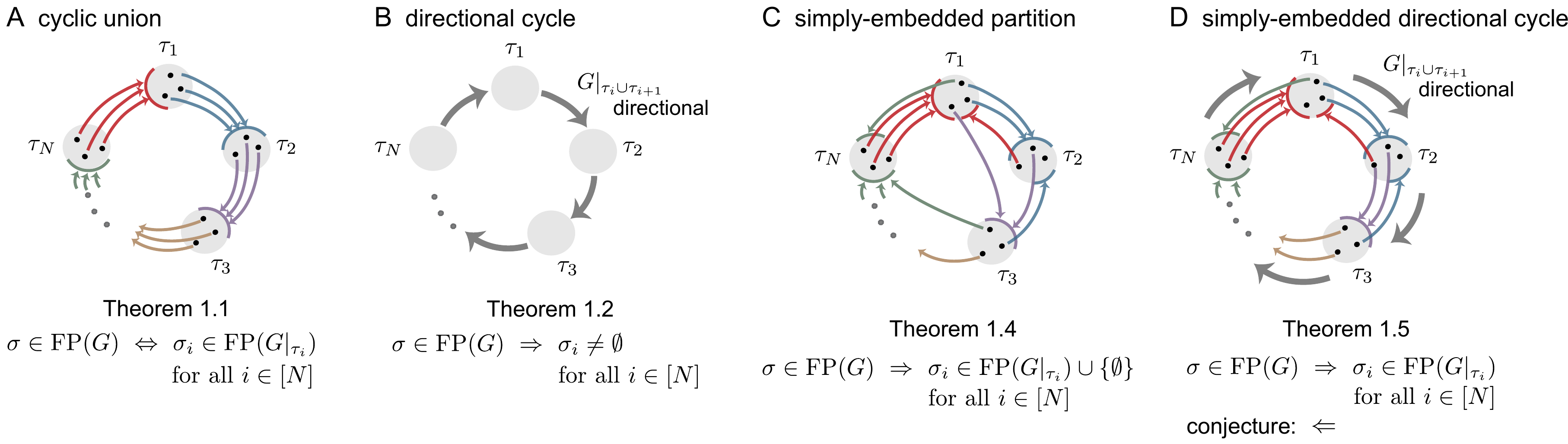}
\vspace{-.33in}
\end{center}
\caption{{\bf Summary of main results.} In each graph, colored edges from a node to a component indicate that the node projects edges out to all the nodes in the receiving component, as needed for a simply-embedded partition.  Thick gray edges indicate directionality of the subgraph $G|_{\tau_{i} \cup \tau_{i+1}}$.  For $\sigma \subseteq [n]$, $\sigma_i \od \sigma \cap \tau_i$.  Theorem numbers match those in \cite{Parmelee2022}.}
\label{fig:cyclic-generalizations-cartoon}
\vspace{-.1in}
\end{figure}

\vspace{-12pt}
\paragraph{Can a single network support multiple types of sequential attractors?\\}

\begin{figure}[!ht]
\begin{center}
\includegraphics[width=0.85\textwidth]{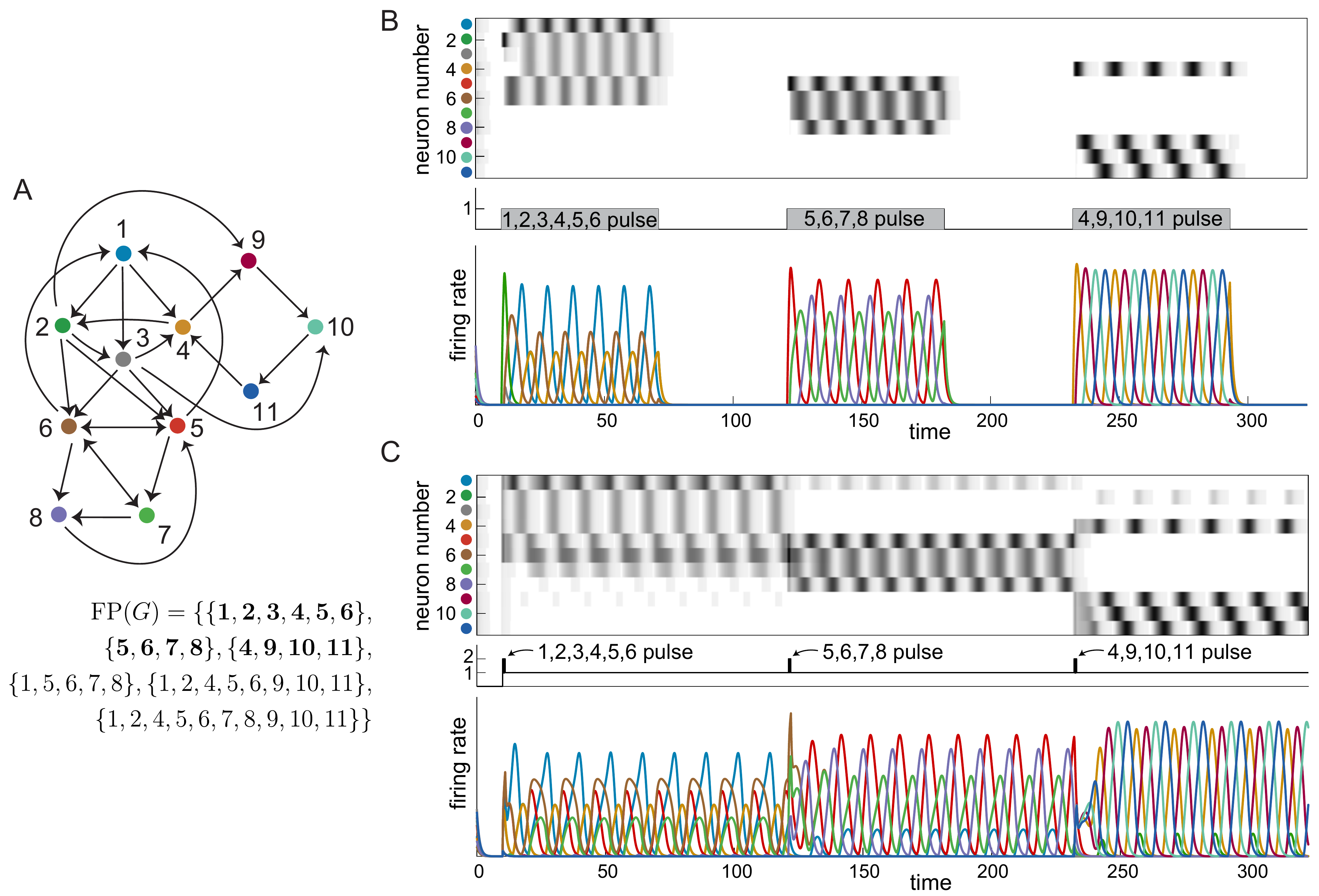}
\vspace{-15pt}
\end{center}
\caption{{\bf Coexistence of three sequential attractors in larger network.} Three embedded attractors are transiently activated to produce sequences matching those of the isolated networks in Figures~\ref{fig:sequences-setup} and \ref{fig:cycu-generalizations}. (A) Three different example graphs are embedded in a single network. $\FP(G)$ is shown, with minimal supports in bold. Notice that the minimal supports correspond exactly to the three subgraphs $G|_{\{1,\dots,6\}},G|_{\{5,6,7,8\}},G|_{\{4,9,10,11\}}$. (B) Long pulses with baseline $\theta = 0$ and pulse values $\theta_i = 1$ activate sequences exactly matching those of the isolated graphs. (C) Short pulses with baseline $\theta = 1$ and pulse values $\theta_i = 2$ activate sequences closely matching those of the isolated graphs.}
\label{fig:attractor-embedding}
\vspace{-10pt}
\end{figure}

The answer is yes! These new architectures can be embedded as subnetworks in a larger network, akin to the example in Figure~\ref{fig:sequences-setup}D, with each subnetwork yielding a different sequential attractor. Figure~\ref{fig:attractor-embedding}A shows a network of eleven neurons made up of three subnetworks on overlapping nodes: $G|_{\{1,\dots,6\}}$ is the cyclic union from Figure~\ref{fig:cycu-generalizations} panel D2, $G|_{\{5,6,7,8\}}$ is the directional cycle from Figure~\ref{fig:cycu-generalizations} panel B2, and $G|_{\{4,9,10,11\}}$ is a 4-cycle (Figure~\ref{fig:sequences-setup}B). The corresponding attractors can be activated by either of two types of external drive. In Figure~\ref{fig:attractor-embedding}B, we chose $\theta = 0$ as a baseline and step function pulses of $\theta_i = 1$, so that only a given subnetwork is receiving positive external drive at a time, and thus only that subnetwork can be active. The pulses activate sequences involving only the neurons in the stimulated subnetwork, which match precisely those in the isolated graphs. In Figure~\ref{fig:attractor-embedding}C, we see that it is also possible for the network to converge to each of the different attractors of these subnetworks even when \emph{all} the neurons receive positive external drive.  Additionally, the network can transition between these attractors simply by providing a short pulse to the relevant subnetwork. In both types of simulations, we see that the network has multiple \emph{coexisting} sequential attractors. Moreover, the minimal fixed points of the network again reflect the subsets of neurons that are high firing in the attractors.

\FloatBarrier 

\vspace{-10pt}
\paragraph{Curious to know more?\\}

The architectures shown in Figure~\ref{fig:cyclic-generalizations-cartoon} all give rise to sequential attractors and have been shown to have well-behaved fixed point structure. In \cite{Parmelee2022}, we further explore these architectures as well as other related network structures, prove theorems, and investigate related questions in depth. We explore sequential attractors of a variety of minimal subnetworks, known as \emph{core motifs}, to find that the directional cycle structure is indeed predictive of the sequential activity, particularly when the directional cycle partition is also simply-embedded. From this analysis, we see that these architectures give insight into the sequences of neural activity that emerge, going beyond the structure of fixed point supports.


\begingroup\footnotesize
\let\section\subsubsection
\makeatletter
\renewcommand\@openbib@code{\itemsep\z@}
\makeatother
\bibliographystyle{unsrt}
\bibliography{CTLN-refs}
\endgroup

\end{document}

%% file: packages-v2.tex
\usepackage{amsmath}
\usepackage{amsthm}
\usepackage{amsfonts}
\usepackage{amssymb}
\usepackage{latexsym} 
\usepackage{epsfig}
\usepackage{graphicx}

\usepackage{calc}  
\usepackage[matrix,tips,graph,curve,web]{xy}

\makeatletter

\def\@maketitle{%
  \newpage
  \null
  \let \footnote \thanks
    {\normalfont\sffamily\bfseries\Large\noindent\@title \par}%
    \vskip 1em%
    {\normalfont\sffamily 
        \noindent
        \@author
        \par}
  \par
  \vskip 4em}
\def\@seccntformat#1{\csname the#1\endcsname{.\ }}
\renewcommand\section{\@startsection {section}{1}{\z@}%
                                   {-3.0ex \@plus -1ex \@minus -.2ex}%
                                   {1.5ex \@plus.2ex}%
                                   {\normalfont\large\bfseries}}
\renewcommand\subsection{\@startsection{subsection}{2}{\z@}%
                                     {-2.75ex\@plus -1ex \@minus -.2ex}%
                                     {1.5ex \@plus .2ex}%
                                   {\normalfont\large}}
\def\fnum@figure{\normalfont\footnotesize\figurename~\thefigure}

\renewcommand\tableofcontents{%
    \section*{\contentsname
        \@mkboth{%
           \MakeUppercase\contentsname}{\MakeUppercase\contentsname}}%
    \@starttoc{toc}%
    }
\renewcommand*\l@part[2]{%
  \ifnum \c@tocdepth >-2\relax
    \addpenalty\@secpenalty
    \addvspace{2.25em \@plus\p@}%
    \begingroup
      \setlength\@tempdima{3em}%
      \parindent \z@ \rightskip \@pnumwidth
      \parfillskip -\@pnumwidth
      {\leavevmode
       \large \bfseries #1\hfil \hb@xt@\@pnumwidth{\hss #2}}\par
       \nobreak
       \if@compatibility
         \global\@nobreaktrue
         \everypar{\global\@nobreakfalse\everypar{}}%
      \fi
    \endgroup
  \fi}
\renewcommand*\l@section[2]{%
  \ifnum \c@tocdepth >\z@
    \addpenalty\@secpenalty
    \addvspace{1.0em \@plus\p@}%
    \setlength\@tempdima{1.5em}%
    \begingroup
      \parindent \z@ \rightskip \@pnumwidth
      \parfillskip -\@pnumwidth
      \leavevmode \sffamily\bfseries
      \advance\leftskip\@tempdima
      \hskip -\leftskip
      #1\nobreak\hfil \nobreak\hb@xt@\@pnumwidth{\hss #2}\par
    \endgroup
  \fi}
\renewcommand*\l@subsection{\sffamily\@dottedtocline{2}{1.5em}{2.3em}}
\renewcommand*\l@subsubsection{\@dottedtocline{3}{3.8em}{3.2em}}
\renewcommand*\l@paragraph{\@dottedtocline{4}{7.0em}{4.1em}}
\renewcommand*\l@subparagraph{\@dottedtocline{5}{10em}{5em}}
\makeatother


\theoremstyle{plain}

\theoremstyle{definition}

  {\begin{list}{}%
    {%
    \settowidth{\labelwidth}{#1}%
    \setlength{\itemindent}{0pt}%
    \setlength{\labelsep}{1em}%
    \setlength{\leftmargin}{\labelwidth+\parindent+\labelsep}%
    \setlength{\itemsep}{0pt}%
    \setlength{\parsep}{.6ex}}}%
  {\end{list}}

  {\begin{list}{}%
    {%
    \settowidth{\labelwidth}{#1}%
    \setlength{\itemindent}{0pt}%
    \setlength{\labelsep}{1em}%
    \setlength{\leftmargin}{\labelwidth+\labelsep}%
    \setlength{\itemsep}{0pt}%
    \setlength{\parsep}{.6ex}}}%
  {\end{list}}

\makeatletter
\newenvironment{subequations*}{
  \begingroup 
  \let\protect\@nx
  \edef\@tempa{\def\@nx\theparentequation{\theequation}}%
  \@xp\endgroup\@tempa
  \setcounter{parentequation}{\value{equation}}%
  \setcounter{equation}{0}%
  \def\theequation{\theparentequation\alph{equation}}%
  \ignorespaces
}{%
  \setcounter{equation}{\value{parentequation}}%
  \global\@ignoretrue
}

\def\d/{/\mspace{-6.0mu}/}